\documentclass[useAMS,usenatbib]{mn2e}

\title{Spectral features of solar wind turbulent plasma}
\author[Dastgeer Shaikh]
  {Dastgeer ~Shaikh \thanks{email: dastgeer.shaikh@uah.edu} and G. P. Zank \\
    Department of Physics and Center for Space Plasma and Aeronomic Research (CSPAR)\\
The University of Alabama in Huntsville,
Huntsville. Alabama, 35899}
\date{Received  on July 8, 2009, Revised on Aug 12, 2009}

\pagerange{\pageref{firstpage}--\pageref{lastpage}} \pubyear{2009}

\def\LaTeX{L\kern-.36em\raise.3ex\hbox{a}\kern-.15em
    T\kern-.1667em\lower.7ex\hbox{E}\kern-.125emX}

\newcommand{\be}{\begin{equation}}
\newcommand{\ee}{\end{equation}}

\newcommand{\eqs}[2]{Eqs. (\ref{#1}) \& (\ref{#2})} 
\newcommand{\Eqs}[2]{Equations  (\ref{#1}) \& (\ref{#2})} 
\newcommand{\Eq}[1]{Equation (\ref{#1})} 
\newcommand{\eq}[1]{Eq. (\ref{#1})} 
\newcommand{\fig}[2]{Figs  (\ref{#1}) \& (\ref{#2})} 
\newcommand{\Fig}[1]{Fig. (\ref{#1})} 
 \newcommand{\eqa}{\begin{eqnarray}}
\newcommand{\eeq}{\end{eqnarray}}

\begin{document}

\label{firstpage}

\maketitle

\begin{abstract}
Spectral properties of a fully compressible solar wind Hall
Magnetohydrodynamic plasma are investigated by means of time dependent
three dimensional Hall MHD simulations. Our simulations, in agreement
with spacecraft data, identify a spectral break in turbulence spectra
at characteristic length-scales associated with electromagnetic
fluctuations that are smaller than the ion gyroradius. In this regime,
our 3D simulations show that turbulent spectral cascades in the
presence of a mean magnetic field follow an omnidirectional
anisotropic inertial range spectrum close to $k^{-7/3}$. The onset of
the spectral break in our simulations can be ascribed to the presence
of nonlinear Hall interactions that modify the spectral cascades. Our
simulations further show that the underlying charachteristic turbulent
fluctuations are spectrally anisotropic, the extent of which depends
critically on the local wavenumber. The fluctuations associated with
length scales smaller than the ion gyroradius are highly compressible
and tend to exhibit a near equipartition in the velocity and magnetic
fields.  Finally, we find that the orientation of velocity and
magnetic field fluctuations critically determine the character of
nonlinear interactions that predominantly govern a Hall MHD plasma,
like the solar wind.

\end{abstract}


\section{Introduction}

The solar wind plasma is predominantly in a turbulent state.
Nonlinear turbulent processes in the magnetized solar wind plasma
plasma fluid yield a multitude of spatial and temporal length-scales
associated with an admixture of waves, fluctuations, structures and
nonlinear turbulent interactions.  In-situ spacecraft measurements
(Matthaeus \& Brown 1988, Goldstein et al 1995, Ghosh et al 1996)
reveal that the solar wind fluctuations, extending over several orders
of magnitude in frequency and wavenumber, when described by power
spectral density (PSD) spectrum, can be divided into {\em three}
distinct regions (Goldstein et al 1995, Leamon et al 1999) depending
on the frequency and wavenumber. The {\em first} region corresponds to
a flatter spectrum, associated with lower frequencies, and it is
consistent with $k^{-1}$ (where $k$ is wavenumber).  {\em Second}
region follows and extends to the ion/proton gyrofrequency, and the
spectral slope has an index ranging from -3/2 to -5/3. This region is
typically characterized as corresponding to fully developed
turbulence, and can be described by the usual incompressible
magnetohydrodynamic (MHD) description.  The turbulent interactions in
this regime are governed entirely by Alfv\'enic cascades.  Spacecraft
observations (Leamon et al 1999, Bale et al. 2005, Alexandrova et al
2007, Sahraoui et al 2009) further reveal that at length scales beyond
the MHD regime, i.e. length scales less than ion gyro radius ($k\rho_i
\gg 1$) and temporal scales greater than the ion cyclotron frequency
$\omega> \omega_{ci}=eB_0/m_ec$, (where $k, \rho_i, \omega_{ci}, e,
B_0, m_e, c$ are respectively characteristic mode, ion gyroradius, ion
cyclotron frequency, electronic charge, mean magnetic field, mass of
electron, and speed of light), the spectrum exhibits a spectral break,
and the spectral index of the solar wind turbulent fluctuations varies
between -2 and -5 (Smith et al 1990, Goldstein et al 1994, Leamon et
al 1999, Bale et al. 2005, Shaikh \& Shukla 2009, Sahraoui et al
2009).  Higher time resolution observations find that at the spectral
break, Alfvenic MHD cascades (Smith et al 1990, Goldstein et al 1994,
Leamon et al 1999) close.  The characteristic modes in this region
appear to evolve typically on timescales associated with dispersive
kinetic Alfvenic fluctuations. 

The onset of the second or the kinetic Alfven inertial range is not
understood. Some suggestions have however been made.  The spectral
break may result from energy transfer processes associated with
possibly kinetic Alfven waves (KAWs) (Hasegawa 1976), electromagnetic
ion-cyclotron-Alfven (EMICA) waves \cite{gary,yoon}, or by
fluctuations described by a Hall MHD (HMHD) plasma model (Alexandrova
et al 2007, 2008; Shaikh \& Shukla 2008, 2008a). Stawicki et al (2001)
argue that Alfv\'en fluctuations are suppressed by proton cyclotron
damping at intermediate wavenumbers so the observed power spectra are
likely to comprise weakly damped dispersive magnetosonic and/or
whistler waves (unlike Alfv\'en waves).  Beinroth \& Neubauer (1981)
and Denskat \& Neubauer (1982) have reported the presence of whistler
waves based on Helios 1 \& 2 observations in this high frequency
regime.  A comprehensive data analysis by Goldstein et al. (1994),
based on correlations of sign of magnetic helicity with direction of
magnetic field, indicates the possibility of the existence of
multiscale waves (Alfv\'enic, whistlers and cyclotron waves) with a
single polarization in the dissipation regime.  Counter intuitively,
in the $\omega < \omega_{ci}$ regime, or Alfv\'enic regime, Howes et
al. (2008) noted the possibility that highly obliquely propagating
KAWs are present (with $\omega \ll \omega_{ci}$) making the
possibility that damping of ion cyclotron waves is responsible for the
spectral breakpoint questionable.  

Fluid (Shaikh \& Shukla 2009) and kinetic (Howes et al. 2008)
simulations, in qualitative agreement with spacecraft data described
as above, have been able to obtain the spectral break point near the
characteristic turbulent length scales that are comparable with ion
inertial length scale ($d_i$). These simulations demonstrated a
Kolmogorov-like $k^{-5/3}$ spectra for the length scales larger than
the ion inertial length scales where MHD is typically a valid
description. By contrast, smaller (than $d_i$) scales are shown to
follow a steeper spectrum that is close to $k^{-7/3}$ (Howes et
al. 2008, Shaikh \& Shukla 2009). Spacecraft data and simulations thus
reveal that migration of turbulent energy proceeds essentially through
different regions in the $k$-space, i.e.  $k^{-1}, k^{-5/3}$ and
$k^{-7/3}$.  Needless to mention that turbulent cascade does not
entirely terminate immediately beyond the $k^{-7/3}$ spectrum. Fluid
and kinetic simulations (Biskamp 1996, Galtier 2006, Galtier \&
Buchlin 2007, Cho and Lazarian 2004, Shaikh \& Zank 2005, Shaikh 2009,
Gary et al. 2008, Saito et al. 2008, Howes et al. 2008) have further
shown that spectral transfer of energy extends even beyond the
$k^{-7/3}$ spectrum and it is governed predominantly by small scale,
high frequency, whistler turbulence. The latter also exhibits a
definite power law.

The physical processes describing MHD, KAW or Hall MHD and whistler
spectra are rich and complex. They differ significantly from others
and continue to pose serious challenges in our understanding of
multiscale solar wind turbulence. One of the major goals of this paper
is to describe the connection between different scales associated with
the MHD, KAW or Hall MHD and whistler spectra. In the following, we
describe extended part of the spectra that are predicted by theory and
simulations.

\begin{figure}
\vspace{200pt}
\begin{center}
\end{center}
\caption{\label{fig1} Schematic of the power spectral density (PSD)
  composite spectrum in the solar wind turbulent plasma as a function
  of frequency (wavenumber). Several distinct regions are identified
  with what are thought to be the dominant energy transfer mechanism
  for that particular region. The nonlinear processes associated with
  the transition from region II (MHD regime) to region III (kinetic or
  Hall MHD regime) are not yet fully understood. The power spectra in
  region III vary from $k^{-2}$ to $k^{-4}$. The boundary of region
  III and IV identifies where electron and ion motions are decoupled.
  Regions IV and V are identified as whistler cascade regimes.}
\end{figure}

\section{extended composite spectra}

Theory and simulations indicate that turbulent fluctuations in the
high frequency and $k\rho_i \gg 1$ (where $\rho_i$ is ion gyro radius)
regime correspond to a regime in which electron motions are decoupled
from the ion motions (Kingsep et al 1990, Biskamp et al 1996, Dastgeer
et al 2000a, Dastgeer et al 2000b, Shaikh \& Zank 2003, Cho \&
Lazarian 2004, Saito et al 2008, Gary et al 2008). Correspondingly,
ions are essentially unmagnetized and can be treated as an immobile
neutralizing background fluid. This regime corresponds to the whistler
wave band of the spectrum and comprises characteristic scales that are
smaller than those that describe MHD, KAW or Hall MHD processes. An
extended composite schematic {\em thus} describing the whistler modes
spectra, in addition to the observed $k^{-1}, k^{-5/3}$ and $k^{-7/3}$
spectra, is shown in \Fig{fig1}. Specifically, regions IV and V in
\Fig{fig1} identify characteristic modes that are relevant for the
description of whistler wave turbulence (Biskamp et al 1996, Shaikh \&
Zank 2005, Shaikh 2009). By contrast, regions I, II and III describe
respectively $k^{-1}, k^{-5/3}$ and $k^{-7/3}$ spectra and are
consistent with the observations (Leamon et al 1999, Bale et al. 2005,
Alexandrova et al 2007, Sahraoui et al 2009).  The boundary of regions
III and IV represents a wavenumber band in spectral space that
corresponds to the decoupling of electron and ion motions. The
wavenumbers above this boundary characterize the onset of whistler
turbulence. The spectral cascades associated with whistler turbulence
are described extensively by Biskamp et al (1996), Dastgeer et al
(2000a), Dastgeer et al (2000b), Shaikh \& Zank (2003), Shaikh \& Zank
(2005), Shaikh (2009a), Shaikh (20009b). Cho \& Lazarian 2004 describe
scale dependent anisotropy that is mediated by whistler waves in the
context of electron MHD plasma. Gary et al. (2008) and Saito et
al. (2008) have reported two-dimensional electromagnetic
particle-in-cell simulations of electron MHD model to demonstrate the
forward cascade of whistler turbulence. Their work show that magnetic
spectra of the cascading fluctuations become more anisotropic with
increasing fluctuation energy. Interestingly, whistler turbulence
associated with longer wavelengths in region IV exhibits a power
spectrum $k^{-7/3}$ that is similar to the short wavelength spectrum
of kinetic Alfv\'en waves (KAW), as shown in region III of
\Fig{fig1}. The underlying physical processes, responsible for the
spectrum differ significantly for KAW and whistler waves. These
differences are discussed in Section 7.

It is to be noted that the Hall MHD description of magnetized plasma
is valid upto region III where characteristic turbulent scales are
smaller than ion inertial length scales ($kd_i > 1$). Beyond this
point, the high frequency motion of plasma is governed predominantly
by the electron motions only. By contrast, ions form static
neutralizing background. Consequently, the ion motions decouple
significantly from the electrons. These aspects of the spectra,
depicted necessarily by regions IV \& V in Fig (1), can be described
adequately by whistler wave model. The Hall MHD models are therefore
not applicable in regions IV, V and beyond. Neither they can describe
kinetic physics associated with the dissipative regime. Since the high
frequency regime (i.e. regions IV \& V) is dominated by the electron
motions, there exists intrinsic length scale corresponding to electron
inerital length scale $d_e = c/\omega_{p_e}$ (where $c$ is the speed
of light and $\omega_{p_e}$ is the electron plasma frequency). The
characteristic turbulent length scales in regions IV \& V are
essentially comparable with $d_e$ and therefore they can describe
scales larger (i.e. $kd_e<1$ in region IV) and smaller (i.e. $kd_e>1$
in region V) than the electron inertial scale. While whistler wave
model can describe nonlinear processes associated with length scales
as small as the electron inertial length scale, they fail to describe
finite electron Larmour radii effects for which a fully kinetic
description of plasma must be sought.

The schematic of solar wind turbulence depicted in \Fig{fig1} raises
numerous unresolved questions. Beside those described above, we do not
understand what leads to the decoupling of ion and electron motions
near the boundary of region III and IV for example. Although the
turbulent spectra are described by similar spectral indices, the
nonlinear processes are fundamentally different in region III and IV.
With this paper, we attempt to identify and understand the processes
that lead to the spectra in III and IV.  It appears that the character
of the nonlinear interaction is determined primarily by the
orientation of turbulent velocity (V) and magnetic (B) field
fluctuations with respect to each other, relative to the mean magnetic
field. The spatially varying angular distribution of perpendicular
velocity and magnetic field fluctuations relative to the mean magnetic
field was predicted theoretically by Boldyrev (2006) and is a
conjecture by Podesta et al (2008).  It nevertheless remains to be
seen how the orientation of turbulent V and B fluctuations govern the
nonlinear spectral transfer of energy in solar wind turbulent
plasma. In the context of MHD turbulence, Servidio et al (2008) show
that the orientation between the velocity and magnetic field
fluctuations plays a critical role in depleting the nonlinear
interactions that lead to the relaxation of MHD turbulence.

Spectral anisotropy is another issue that is not yet properly
understood for characteristic modes with scales greater than $kd_i
\sim 1$ in Hall MHD plasma. In the case of MHD turbulence, the
presence of a mean magnetic field leads to the asymmetric transfer of
spectral energy along and across the mean magnetic field \cite{sheb}.
An excellent analysis of anisotropic cascades in the framework of Hall
MHD is presented by Ghosh \& Goldstein (1997).  They describe how the
degree of turbulent anisotropy in a Hall MHD plasma varies as the
plasma beta (the ratio of magnetic to pressure energy) changes from
greater than to less than 1. A more general characterization of
spectral anisotropy in the vicinity of modes $kd_i \sim 1$ is still
unclear. Specifically, the scale dependence of the turbulent
anisotropy in the KAW regime and it's connection with that in the
whistler regime is not yet established.  Spectral anisotropy in the
whistler regime was independently investigated by Shaikh \& Zank
(2003), Dastgeer et al (2000a, b). In this paper, we relate these
results to those associated with anisotropic KAW modes.

To address the issues described above, we use time dependent, fully
compressible three dimensional simulations of Hall MHD plasma in a
triply periodic domain.  This represents a local or regional volume of
the solar wind plasma.  Note that the dynamics of length-scales
associated with region III, i.e. corresponding to the KAW modes,
cannot be described by the usual MHD models since we are interested in
characteristic frequencies smaller than an ion gyro frequency.  At 1
AU, ion inertial length scales are smaller than ion gyro radii in the
solar wind \cite{Goldstein1995}.  Plasma effects due to finite Larmor
radii can readily be incorporated in MHD models by introducing Hall
terms to accommodate ion gyro scales up to scales as small as ion
inertial length scales.  In section 3, we describe the underlying Hall
MHD model along with the interinsic assumptions and discuss the Hall
MHD linear dispersion relation. Section 4 addresses our 3D fluid
simulations of the nonlinear Hall MHD equations. Our results suggest
that the secondary inertial range spectrum has a form that is close to
$k^{-7/3}$ above the spectral break in the solar wind plasm, and
mediated by the Hall terms in the short wavelength (in comparison with
the ion skin-depth) KAW regime.  Anisotropy in spectral cascade
behavior is explored in section 5. We find that long length scale
fluctuations in the $kd_i>1$ KAW regime exhibit a more anisotropic
energy cascade compared to smaller scales.  The dynamical alignment
and angular distribution of turbulent velocity and magnetic field
fluctuations is described in section 6. We find that characteristic
turbulent flucutations in the $kd_i>1$ regime relax towards a state of
orthogonality such that the majority of turbulent scales contain
fluctuations in which the velocity and magnetic fields are nearly
orthogonal , i.e. ${\bf V} \perp {\bf B}$.  Section 7 compares the KAW
and whistler spectra. Finally, in section 8 we provide a summary and
conclusions.

\section{Hall MHD Simulation Model}
Our 3D simulations are based on a two fluid nonlinear Hall MHD plasma
model. The model assumes that the electrons are inertial-less, while
the ions are inertial \cite{mahajan}. Hence, the electrons and ions
have a differential drift, unlike the one fluid MHD model for which
the electron and ion flow velocities are identical. Hall MHD
description of magnetized plasma has previously been employed by a
number of workers to investigate wave and turbulence processes in the
context of solar wind plasma. In an excellent work, Sahraoui et
al. (2007) extended the ordinary MHD system to include spatial scales
down to the ion skin depth or frequencies comparable to the ion
gyrofrequency in an incompressible limit. They further analyzed the
differences in the incompressible Hall MHD and MHD models within the
frame work of linear modes, their dispersion and
polarizations. Galtier (2006) developed a wave turbulence theory in
the context of an incompressible Hall MHD system to examine the
steepening of the magnetic fluctuation power law spectra in the solar
wind plasma. Furthermore, Galtier and Buchlin (2007) have developed 3D
dispersive Hall magnetohydrodynamics simulations within the paradigm
of a highly turbulent shell model and demonstrated that the
large-scale magnetic fluctuations are characterized by a
$k^{-5/3}$-type spectrum that steepens at scales smaller than the ion
inertial length $d_i$ to $k^{-7/3}$.

Here we start from the
electron momentum equation, 
\be m_e n_e \left(
\frac{\partial}{\partial t} + {\bf V}_e \cdot \nabla \right) {\bf V}_e
= - \nabla P_e - en_e \left({\bf E} + \frac{1}{c} {\bf V}_e \times
{\bf B} \right) \ee where $m_e, n_e, {\bf V}_e, P_e$ are respectively
mass, density, velocity and pressure of electrons, and ${\bf E}$ and
${\bf B}$ are the electric and magnetic fields.  In the presence of
low-frequency (compared with the electron gyrofrequency)
electromagnetic fields, the electric force acting on the inertial-less
electrons is balanced by the electron Lorentz force and the pressure
gradient.  This yields a more general form of Ohm's law than is
typically used in the MHD description and is described as Hall MHD.
We assume a quasineutral solar wind plasma where the density of
electrons ($n_e$) and ions ($n_i$) is nearly equal such that $n_e
\approx n_i = n$. Thus in the inertial-less (Ohm's law) electron
limit, the electron momentum equation yields the electric field as \be
{\bf E} = - \frac{1}{ne}\nabla P_e - \frac{1}{c} {\bf V}_e \times {\bf
  B} \ee The electric field arising from separation introduced by the
inertial-less electron momentum equation can be substituted into the
ion momentum equation \be m_i n_i \left( \frac{\partial}{\partial t} +
{\bf V}_i \cdot \nabla \right) {\bf V}_i = - \nabla P_i - en_i
\left({\bf E} + \frac{1}{c} {\bf V}_i \times {\bf B} \right), \ee
where $m_e, n_e, {\bf V}_e, P_e$ are mass, density, velocity and ion
pressure respectively, which then yields \be \rho \left(
\frac{\partial}{\partial t} + {\bf V}_i \cdot \nabla \right) {\bf V}_i
= - \nabla P + \frac{1}{c} {\bf J} \times {\bf B}.  \ee Here we used
$\rho = m_i n$ (solar wind plasma mass density), ${\bf J} = ne ({\bf
  V}_i-{\bf V}_e) $ (plasma current) and $P=P_e+P_i$ (total plasma
pressure). We next substitute the electric field into Faraday's law
which yields \be \frac{\partial {\bf B}}{\partial t} = -c\nabla \times
{\bf E}=\nabla\times ( {\bf V}_e \times {\bf B}).  \ee On eliminating
the electron fluid velocity from the current equation i.e. ${\bf V}_e
={\bf V}_i -{\bf J}/ne $, we obtain \be \frac{\partial {\bf
    B}}{\partial t} = \nabla \times \left({\bf V}_i \times {\bf B} -
\frac{{\bf J} \times {\bf B}}{ne} \right).  \ee

In the inertial-less electron ($m_e \rightarrow 0$) limit the electron
fluid does not influence the momentum of solar wind plasma directly
except through the current. Since the electron fluid contributes to
the electric field, plasma currents and the magnetic field are
affected by electron oscillations. The combination of electron
dynamics and ion motions distinguishes the Hall MHD model from its
single fluid MHD counterpart. Thanks to the inclusion of electron
dynamics, Hall MHD can describe solar wind plasma fluctuations that
are associated with a finite ion Larmor radius and thus a
characteristic plasma frequency is $\omega > \omega_{ci}$.  Because
Hall MHD contains both ion and electron effects, there is a regime at
which the dominance of one set of plasma fluctuations crosses over to
be dominated by the other. This therefore introduces naturally an
intrinsic scale length/timescale (frequency) that separates ion
dominated behavior in the plasma from electron dominated.  It is the
Hall term corresponding to the ${\bf J} \times {\bf B}$ term in
Faraday's law that is primarily responsible for decoupling electron
and ion motion on ion inertial length and ion cyclotron time scales
(and introducing an intrinsic length scale). It is this feature that
makes Hall MHD useful in describing dissipative solar wind processes
when single fluid MHD is not applicable (the MHD model breaks down at
$\omega > \omega_{ci}$). Hall MHD allows us to study inertial range
cascades beyond $\omega > \omega_{ci}$, and can be extended to study
dissipative heating processes where ion cyclotron waves are damped.
The extreme limit of a fluid modeling applied to solar wind processes
(even beyond the limit of the Hall MHD regime) is the use of an
electron MHD model in which high frequency electron dynamics is
treated by assuming stationary ions that act to neutralize the plasma
background.  We discuss this regime at the end of this paper.

The quasi-neutral plasma density ($\rho$), the velocity (${\bf V}$),
the magnetic field (${\bf B}$) and the plasma pressure ($P=P_e+P_i$)
can be described by the nonlinear Hall MHD equations. For the purpose
of numerical simulations, it is convenient to express Hall MHD in the
conservative form,

\be
\label{hmhd}
 \frac{\partial {\bf F}}{\partial t} + \nabla \cdot {\bf Q}={\cal Q},
\ee
where,
\[{\bf F}=
\left[ 
\begin{array}{c}
\rho  \\
\rho {\bf V}  \\
{\bf B} \\
e
  \end{array}
\right], 
{\bf Q}=
\left[ 
\begin{array}{c}
\rho {\bf V}  \\
\rho {\bf V} {\bf V}+ \frac{P}{\gamma-1}\bar{I}+\frac{B^2}{8\pi} \bar{I}-{\bf B}{\bf B} \\
{\bf V}{\bf B} -{\bf B}{\bf V} - d_i ({\bf J}{\bf B} -{\bf B}{\bf J})\\
{\bf W}
  \end{array}
\right]; \]
\[ {\cal Q}=
\left[ 
\begin{array}{c}
0  \\
{\bf f}_M({\bf r},t) +\mu \nabla^2 {\bf V}+\eta \nabla (\nabla\cdot {\bf V})  \\
\eta \nabla^2 {\bf B}  \\
0
  \end{array}
\right],
\] 
and
\[e=\frac{1}{2}\rho V^2 + \frac{P}{(\gamma-1)}+\frac{B^2}{8\pi},\]
\[{\bf W}
= \frac{1}{2} \rho V^2{\bf V} + \frac{\gamma P {\bf V}}{(\gamma-1)} 
+ \frac{c}{4\pi}{\bf E} \times {\bf B}. \]
The suffix $i$ is dropped from the ion fluid velocity ${\bf V}_i$ in the
conservative form.
  
The dynamical variables are functions of three space coordinates and
time, i.e. $(x,y,z,t)$ and are normalized by typical length $\ell_0$
and time $t_0 = \ell_0/V_A$ in our simulations, $V_A=B_0/(4\pi
\rho_0)^{1/2}$ the Alfv\'en speed, such that
$\bar{\nabla}=\ell_0{\nabla}, \partial/\partial
\bar{t}=t_0\partial/\partial t, \bar{\bf V}={\bf V}/V_A,\bar{\bf B}
={\bf B}/V_A(4\pi \rho_0)^{1/2}, \bar{\bf E} ={\bf E}/V_A(4\pi
\rho_0)^{1/2}, \bar{P}=P/\rho_0V_A^2, \bar{\rho}=\rho/\rho_0,
\bar{c}=c/V_A$.  $\bar{I}$ is the unit tensor. The parameters $\mu$
and $\eta$ represent the ion-electron viscous drag and magnetic field
diffusivity, respectively. While the viscous drag modifies the
dissipation in the plasma momentum in a nonlinear fashion, the
magnetic diffusivity damps small scale magnetic field fluctuations
linearly.  The dimensionless parameter $\bar{d}_i (=d_i/\ell_0$, where
$d_i =c/\omega_{pi}$ is the ion skin depth and $\omega_{pi}$ is the
ion plasma frequency) in Faraday's law identifies the Hall effect.
The ion skin depth is a natural or an intrinsic length scale
representative of the Hall MHD plasma model, and the Hall term plays
an important role for high-frequency fluctuations with $k d_i \geq 1$.
It turns out that the Hall physics dominates the magnetoplasma
dynamics when $(1/\rho){\bf J} \times {\bf B} > {\bf V} \times {\bf
  B}$ in the magnetic field induction equation.  Finally, our
nonlinear Hall MHD model also includes the full energy equation,
rather than an adiabatic equation of state connecting the plasma
pressure and the density.  The use of the energy equation enables us
to follow self-consistently the evolution of turbulent plasma heating
resulting from the nonlinear cascading of energy.

The rhs in the momentum equation corresponds to a forcing function
(${\bf f}_M({\bf r},t)$) that drives the plasma momentum at large
length scales in our simulation.  With the help of this function, we
introduce energy in large scale eddies to sustain the magnetized
turbulent interactions. In the absence of forcing, the turbulence
decays freely.  While the driving term modifies the plasma momentum,
we conserve density (since we neglect photoionization and
recombination). The large-scale random driving of turbulence can
correspond to external forces or instabilities, for example fast and
slow streams, merged interaction regions etc, in the solar wind, (or
even supernova explosions and stellar winds in the ISM, etc).

The linear Hall MHD equations admit the dispersion
relation \cite{brodin}, 
\[(\omega^ 2-k_z^ 2V_A^ 2)D_m (\omega, {\bf k})
=\frac{\omega^ 2}{\omega_{ci}^ 2}(\omega^ 2-k^ 2V_s^ 2) k_z^ 2k ^2V_A^ 4,\]
which has been written in a form to clearly identify the relationship
of MHD to Hall MHD. The expression
\[D_m(\omega, {\bf k}) =\omega^ 4 -\omega^ 2
k^ 2 (V_A^ 2+V_s^ 2)+k_z^ 2k^ 2 V_A^ 2 V_s^ 2\] 
is the MHD dispersion relation for fast and slow magnetosonic
modes. Here $\omega$ is the wave frequency, ${\bf k} (={\bf k}_\perp +
k_z \hat {\bf z}$) the wave vector, the subscripts $\perp$ and $z$
represent components across and along the external magnetic fields
$B_0 \hat {\bf z}$, and $V_A$ and $V_s$ are the Alfv\'en and sound
speeds respectively. Besides the fast and slow magnetosonic modes,
KAWs, EMICA waves, and electron whistlers are identified by the
remaining three roots of the above dispersion relation.  Warm Hall MHD
plasma thus supports a great variety of waves of different wavelengths
(comparable to the ion skin depth associated with the Hall drift, the
ion sound gyroradius associated with the electron pressure and the
perpendicular ion inertia/ion polarization drift).

\begin{figure}
\vspace{200pt}
\caption{Scaling of our three dimensional Hall MHD code with respect
  to number of processors. The spectral resolution is fixed at $60^3$
  in a cubic box of volume $\pi^3$. By increasing the number of shared
  memory processors, the total computational time (measured in CPU
  unit time; real time) to achieve steady state turbulent interactions
  can be reduced significantly.  Our code clearly scales efficiently
  with an increasing number of processors. }
\label{scaling_fig}
\end{figure}

\section{Nonlinear simulation results}
To study the evolution of turbulent cascades in the solar wind
turbulent plasma, we have developed a fully 3D compressible Hall MHD
code.  The three dimensional computations are numerically
expensive. But, with the advent of high speed vector and parallel
distributed memory clusters, and efficient numerical algorithms such
as those designed for Message Passing Interface (MPI) libraries, it is
now possible to perform magnetofluid turbulence studies at
substantially higher resolutions. Based on MPI libraries, we have
developed a three dimensional, time dependent, compressible,
non-adiabatic, driven and fully parallelized Hall magnetohydrodynamic
(MHD) nonlinear code that runs efficiently on both distributed memory
clusters like distributed-memory supercomputers or shared memory
parallel computers. This allows us to achieve a very high resolution
in Fourier spectral space. The code is scalable and transportable to
different cluster machines.  The spatial descretization employs a
conservative and parallelized descretization of Fourier harmonics, and
the temporal integration uses a Runge Kutta (RK) 4th order method.
The boundary conditions are periodic along the $x,y$ and $z$
directions in the local rectangular region of the solar wind
plasma. The MHD counterpart of this code was used in several other
studies by us, see e.g. \cite{dastgeer06,dastgeer07,dastgeer09}.

Our code treats the solar wind plasma fluctuations as statistically
isotropic, locally anisotropic, homogeneous and random. Such a
representation is further consistent with ACE spacecraft measurements
(Smith et al 2006). The numerical algorithm accurately describes the
physical variables (density, temperature, magnetic field, velocity
field, pressure etc) in our code and they are also less
dissipative. Because of the latter, nonlinear mode coupling
interactions preserve ideal rugged invariants of fluid flows, unlike
finite difference or finite volume methods. The conservation of ideal
invariants (energy, enstrophy, magnetic potential, helicity) in
inertial range turbulence is an extremely important feature because
these quantities describe the cascade of energy in the inertial
regime, where turbulence is, in principle, free from large-scale
forcing as well as small scale dissipation. Damping of plasma
fluctuations may nonetheless occur as a result of intrinsic non-ideal
effects such those introduced by the finite Larmor radius.

To test the scalability of our code, we have performed simulations by
increasing the number of processors for a fixed number of modes. Our
scaling results are depicted in \Fig{scaling_fig}. Clearly, by
increasing the number of shared memory processors, the computational
time to achieve steady state turbulent interactions can be reduced
significantly.

The turbulent fluctuations are initialized by using a uniform
isotropic random spectral distribution of Fourier modes concentrated
in a smaller band of lower wavenumbers ($k<0.1~k_{max}$). While
spectral amplitudes of the fluctuations are random for each Fourier
coefficient, it follows a certain initial spectral distribution
proportional to $k^{-\alpha}$, where $\alpha$ is the initial spectral
index.  The spectral distribution set up in this manner initializes
random scale turbulent fluctuations.  We note that a constant magnetic
field is included along the $z$ direction (i.e. ${\bf B}_0 =B_0 \hat
{\bf z}$) to accommodate the large scale (or the background solar
wind) magnetic field.  The plasma beta, ratio of plasma pressure to
magnetic field energy, is close to unity in our simulation $\beta
\simeq 1$, as typically observed in solar wind (Goldstein et al 1995;
Smith et al 2006).  Turbulent fluctuations in our 3D Hall MHD
simulations are driven either at the lowest Fourier modes or evolve
freely under the influence of the self-consistent dynamics described
by the set of \Eq{hmhd}.  The inertial range spectral cascades in
either cases lead to nearly identical inertial-range turbulent
spectra. We have further carried out 3D simulations for a range of
various parameters and spectral distributions to ensure the validity,
as well as the consistency, of our codes and the physical results.
The simulation parameters are; spectral resolution is $128^3$,
$\eta=\mu=10^{-3}, \beta=1.0, kd_i \sim 0.1-10, L_x=L_y=L_z=2\pi$.  The
nonlinear coupling of velocity and magnetic field fluctuations, amidst
density perturbations, excites high-frequency and short wavelength (by
the $\omega/\omega_{ci}$ effect) compressional dispersive KAWs.  While
the nonlinear mode coupling interactions must influence local
turbulent fluctuations in the inertial range, their role in the
spectral energy cascade is not yet fully understood.  We come back to
this issue below.

\begin{figure}
\vspace{200pt}
\begin{center}
\end{center}
\vbox{Landscape/two column figure to go here}
\caption{Inertial range turbulent spectra for magnetic and velocity
  field fluctuations. The fluctuations closely follow respectively
  $k^{-5/3}$ and $k^{-7/3}$ scalings in the $kd_i<1$ and $kd_i>1$ KAW
  regimes.  $d_i=0.05$ and $1.0$ respectively in the $kd_i<1$ and
  $kd_i>1$ regimes. The dash-dot straight lines correspond to a
  $k^{-5/3}$ and a $k^{-7/3}$ power law. }
\label{spectra}
\end{figure}

The nonlinear spectral cascade in the modified KAW regime leads to a
secondary inertial range in the vicinity of $kd_i \simeq 1$, where the
turbulent magnetic and velocity fluctuations form spectra close to
$k^{-7/3}$. This is displayed in \Fig{spectra}, which also shows that
for length scales larger than the ion thermal gyroradius lead, an MHD
inertial range spectrum close to $k^{-5/3}$ is formed.  Figure
(\ref{spectra}) results from averaging the spectra over ten simulation
runs that are initialized with different random numbers. The spectra
obtained in each simulation further consist of 20 different spectra
that are averaged over the simulation period $(15 -20)~l_0/V_A$. The
resultant spectrum is further processed using a higher order
polynomial fit leading thereby to a well-behaved omnidirectional power
law.  The characteristic turbulent spectrum in the KAW regime is
steeper than that of the MHD inertial range.  Identifying the onset of
the secondary inertial range has been the subject of some debate
because of the presence of multiple processes in the KAW regime that
can mediate the spectral transfer of energy. These processes include,
for instance, the dispersion and damping of EMICA waves, turbulent
dissipation, etc. In the context of our 3D Hall MHD simulations, the
observed $k^{-7/3}$ scaling in the turbulent plasma can be understood
from effect of the Hall term on the energy cascade process.  The
time-scale associated with Hall MHD is $\tau_{\rm H}$, which is
smaller than the MHD characteristic time scale $\tau_{\rm MHD}$.
Nonlinear cascades in MHD turbulence are typically governed by
\[\tau_{\rm MHD} \sim \frac{1}{k v_k},\]
where $v_k=|{\bf V}({\bf k},t)|$ is the velocity field in $k$-space.
By contrast, spectral transfer of turbulent energy in the Hall MHD plasma
has a typical timescale of
\[\tau_{\rm H} \sim \frac{1}{k^2 B_k}.\]
 The energy transfer rate in the
KAW regime is, therefore, 
\be
\varepsilon \sim \frac{v_k^2}{\tau_{\rm H}}.
\label{hmhd-spectrum}
\ee 
On assuming a turbulent equipartition relation
\cite{alexandrova2007,alexandrova2008} between the velocity and
magnetic fields $B_k^2 \sim v_k^2$, and using $\tau_{\rm H}$, the
energy transfer rate, \eq{hmhd-spectrum} reduces to
\[\varepsilon \sim k^2 B_k^3.\]  

The inertial range modes in our 3D Hall MHD simulations exhibit near
equipartition, a result that is consistent with other work
\cite{alexandrova2007,alexandrova2008}.  This is shown in
\Fig{equipartition_time}.  It is noteworthy that equipartition was
assumed by Alexandrova et al (2007) and Alexandrova et al (2008) and
is not directly observed in the solar wind plasma in the $kd_i>1$
regime.  Our simulation results shown in \Fig{equipartition_time} thus
provide theoretical support to arguments that invoke turbulent
equipartition in the MHD/KAW range the
\cite{alexandrova2007,alexandrova2008}.  \Fig{equipartition_time}
shows turbulent equipartition where spectrally averaged (i.e. averaged
over entire $k$-spectrum) fluctuations exhibit turbulent equipartition
during the nonlinear evolution. In \Fig{equipartition_time}, the ratio
of magnetic and velocity field energy remains close to unity
i.e. $|B_k|^2/|v_k|^2 \sim 1$. \Fig{delBdelV_k} describes spectral
distribution of $|\delta v|^2/|\delta B|^2$ as a function of $k$. This
figure also shows that the ratio $|\delta v|^2/|\delta B|^2$ stays
more or less close to the unity for the inertial range turbulent
scales ($kd_i \ge 1$) in our simulations.

In the $kd_i<1$ Alfv\'enic regime, turbulent equipartition in the
solar wind plasma remains a subject of debate
\cite{podesta,TuMarsch,yokoi2006,yokoi2007}.  Turbulent equipartition
observed in the solar wind plasma in the $kd_i<1$ Alfv\'enic regime is
not exact and $v_k^2/B_k^2 \le 1$.  A precise mechanism leading to the
non-exact turbulent equipartition is unclear in the solar wind plasma
(in the $kd_i<1$ Alfv\'enic regime).  The deviation from equipartition
may result from inhomogeneity of the magnetic field \cite{yokoi2006},
possible differences in the dissipation of kinetic and magnetic
energies \cite{yokoi2007}, or compressibility of the density
fluctuations \cite{podesta}. Our objective however is not to address
the mechanism that lead to unequal turbulent equipartition in the
solar wind plasma in the $kd_i<1$ Alfv\'enic regime. The readers can
refer to the work of \cite{podesta,TuMarsch,yokoi2006,yokoi2007} for a
more detailed analysis of turbulent equipartition in the $kd_i<1$
Alfv\'enic regime in the solar wind plasma. We simply use the
equipartition relationship here to derive Kolmogorov-like energy
spectrum in Hall MHD plasma.

To derive the expression for the energy spectrum describing the energy
cascade in Hall MHD plasma, we apply Kolmogorov's phenomenology
\cite{kol,krai,iros} in which the energy cascade in the inertial range
is local and  depends on Fourier modes and the energy dissipation
rates \cite{kol,krai,iros}. We obtain
\[k^{-1}B_k^2 \sim (B_k^3 k^2)^{\alpha} k^{\beta}.\]
On equating indices, we obtain $\alpha = 2/3$ and $\beta = -7/3$.
This results in an energy spectrum of the form
\[E_k \propto k^{-7/3},\] 
which is consistent with our 3D Hall MHD simulations [see
  \Fig{spectra}] and solar wind observations (Goldstein et al 1995;
Leamon et al 1999, Bale et al. 2005, Sahraoui et al 2009, Alexandrova
et al 2007, 2008).  On the other hand, the use of $\tau_{\rm MHD}$ in
estimating the energy dissipation rates recovers the $k^{-5/3}$
inertial range MHD spectrum. This suggests that the Hall effect may be
responsible for the spectral steepening in the solar wind plasma
fluctuations in the $kd_i>1$ regime.

It is worth mentioning that \Fig{equipartition_time} represents
spectral average of the entire mode spectrum.  On the other hand,
spectra of $|\delta B(k)|^2/|\delta V(k)|^2$ and $| B(k)|^2/| V(k)|^2$
in steady state show that turbulent equipartition in Hall MHD deviates
for higher $k$'s (Shaikh \& Shukla 2009). Interestingly, nonlinear
fluid simulations in whistler wave regime (Dastgeer et al 2000a,
Shaikh 2009) corresponding to region IV in Fig (1) show that magnetic
and velocity fields in whistler modes tend to establish turbulent
equipartition. This is essentially a regime where the characteristic
length scales are bigger than electron inertial length, i.e.
$kd_e<1$.  In this regime, dispersive wave effects dominate the
cascade processes by establishing turbulent equipartion amongst
modes. In the $kd_e>1$ regime (region V in Fig (1)), Dastgeer et al
(2000b) find that turbulent equipartion does not hold for short scales
in EMHD turbulence. The latter behaves like hydrdynamic regime of
fluid turbulence where total energy is dominated entirely by kinetic
energy, whereas magnetic energy becomes sub-dominant.

\begin{figure}
\vspace{200pt}
\begin{center}
\end{center}
\caption{Volume averaged spectra of velocity and magnetic field
fluctuations exhibit turbulent equipartition as a function of time. }
\label{equipartition_time}
\end{figure}

\begin{figure}
\vspace{200pt}
\begin{center}
\end{center}
\caption{
Spectral distribution of $|\delta v|^2/|\delta B|^2$ as
a function of $k$. Figure shows that the ratio $|\delta
v|^2/|\delta B|^2$ stays more or less close to the unity. }
\label{delBdelV_k}
\end{figure}

\section{Anisotropic cascades}

Here we discuss the degree of anisotropy in the turbulence, introduced
by the presence of a large scale magnetic field in 3D Hall MHD. A mean
magnetic field in the $z$-direction is included in our simulation,
this meant to mimic the large scale solar wind background magnetic
field. The small scale fluctuations are influenced by the presence of
the large scale mean magnetic field and tend to evolve in an
anisotropic manner in the sense that the turbulent cascade along and
across the mean magnetic field behave differently. In 3D turbulence,
fluctuations in a plane orthogonal to the mean magnetic field remains
isotropic, whereas those aligned in the direction of the external
magnetic field are affected. The latter is establised for $\beta<1$
and $\beta \simeq 1$ by Zank \& Matthaeus (1990, 1993) in the context
of a nearly incompressible MHD theory.  Thus, turbulent anisotropy in
the inertial range spectrum corresponds to the preferential transfer
of spectral energy that feeds perpendicular modes $k_\perp$ while the
parallel ($k_\parallel$) cascades are suppressed. The anisotropy in
the initially isotropic turbulent spectrum is triggered essentially by
background anisotropic gradients that nonlinearly migrate spectral
energy in a particular direction. To measure the degree of anisotropic
cascades, we employ the following diagnostics to monitor the evolution
of the $k_\perp$ mode in time. The averaged $k_\perp$ mode is
determined by averaging over the entire turbulent spectrum weighted by
$k_\perp$, thus
\[ \langle k_\perp(t)\rangle = \sqrt{\frac{\sum_k |k_\perp Q(k,t)|^2}{\sum_k |Q(k,t)|^2}}. \]
Here $\langle \cdots \rangle$ represents an average over the entire
Fourier spectrum, $ k_\perp=\sqrt{k_x^2+k_y^2}$ and $Q$ represents any
of $B, V, \rho$, $\nabla \times {\bf B}$ and $\nabla \times {\bf V}$.
Similarly, the evolution of the $k_\parallel$ mode is determined by
the following relation,
\[ \langle k_\parallel(t) \rangle= \sqrt{\frac{\sum_k |k_\parallel Q(k,t)|^2}{\sum_k |Q(k,t)|^2}}. \]
It is clear from these expressions that the $\langle
k_\perp(t)\rangle$ and $\langle k_\parallel(t) \rangle$ modes exhibit
isotropy when $\langle k_\perp(t)\rangle \simeq \langle k_\parallel(t)
\rangle$. Any deviation from this equality corresponds to spectral
anisotropy. We follow the evolution of the $\langle k_\perp(t)\rangle$
and $\langle k_\parallel(t) \rangle$ modes in our simulations.  Our
simulation results describing the evolution of $\langle k_\perp
\rangle$ and $\langle k_\parallel(t) \rangle$ modes are shown in
\Fig{aniso}. It is evident from \Fig{aniso} that the initially
isotropic modes $\langle k_\perp(t)\rangle \simeq \langle
k_\parallel(t) \rangle$ gradually evolve towards a highly anisotropic
state in that spectral transfer preferentially occurs in the $\langle
k_\perp(t)\rangle$ mode, and is suppressed in $\langle k_\parallel(t)
\rangle$ mode. Consequently, spectral transfer in the $\langle
k_\perp(t)\rangle$ mode dominates the nonlinear evolution of
fluctuations in Hall MHD, and mode structures become elongated along
the mean magnetic field or $z$-direction. Hence nonlinear interactions
led by the nonlinear terms in the presence of background gradients
lead to anisotropic turbulent cascades in the inertial range turbulent
spectra.

\begin{figure}
\vspace{200pt}
\begin{center}
\end{center}
\caption{\label{fig4} Evolution of $\langle k_\perp(t)\rangle $ and $
  \langle k_\parallel(t) \rangle $ as a measure of anisotropy in Hall
  MHD. Initially, $ \langle k_\perp(t=0) \rangle =
  k_\parallel(t=0)$. As time evolves, anisotropy in spectral transfer
  progressively develops such that $\langle k_\perp(t)\rangle >
  \langle k_\parallel(t) \rangle $. Hence the presence of a large
  scale magnetic field suppresses the turbulent cascade along the
  direction of the field, while the perpendicular cascade remains
  nearly unaffected.}
\label{aniso}
\end{figure}

\begin{figure}
\vspace{200pt}
\begin{center}
\end{center}
\caption{\label{modes} Spectrum of $ k_\perp$ and $ k_\parallel$ as a
  function of $k$.  The large scale inertial range turbulent
  fluctuations are more anisotropic as compared to the smaller ones.}
\label{aniso2}
\end{figure}

While the spectral transfer of energy differs in the $k_\perp$ and
$k_\parallel$ modes, the 3D volume averaged turbulent spectrum follows
a $k^{-7/3}$ power law, as shown in \Fig{spectra}.  The steepness of
the observed spectrum can be ascribed to the coexistence of partially
anisotropic flows and turbulent fluctuations in steady state Hall MHD
turbulence.  \Fig{aniso} illustrates anisotropy corresponding to an
averaged $k$ mode but the anisotropy exhibited by the small and large
scale ${\bf B}$ and ${\bf v}$ fluctuations is not distinctively clear,
nor is the degree of anisotropy in ${\bf B}$ and ${\bf v}$ fields
clear from \Fig{aniso}.  The scale dependece of turbulent anisotropy
is described in \Fig{aniso2}. \Fig{aniso2} shows discrepancy in
$k_\perp$ and $k_\parallel$ is prominent at the smaller $k$'s.  This
essentially means that the large scale turbulent fluctuations are more
anisotropic than the smaller ones in a regime where characteristic
length scales are smaller than $d_i$ i.e. $kd_i > 1$.  It further
appears from \Fig{aniso2} that the smaller scales in the $kd_i > 1$
are virtually {\em unaffected} by anisotropic kinetic Alfv\'en waves
that propagate along the externally imposed mean magnetic field $B_0$.
Turbulent fluctuations with small characteristic scales in the $kd_i >
1$ regime of Hall MHD are not affected by the mean magnetic field or
kinetic Alfv\'en waves. This leads us to conjecture that small scale
turbulence in the $kd_i > 1$ regime behaves essentially
hydrodynamically i.e.  as eddies independent of the mean magnetic
field or collisionless magnetized waves.  Thus large and smaller
turbulent length scales evolve differently in the $kd_i > 1$ regime of
Hall MHD turbulence.

\begin{figure}
\vspace{200pt}
\begin{center}
\end{center}
\caption{ Evolution of the angular anisotropic distribution in Hall
  MHD turbulent plasma fluctuations. Shown in the figure is the
  angular evolution of anisotropy associated with the rotational
  velocity and magnetic field fluctuations. The initial angular
  distribution is identical.  As the simulation progresses, small
  scale velocity field fluctuations tend to become more anisotropic
  than their magnetic field counterpart.  The implication is that the
  smaller scale magnetic field fluctuations are less anisotropic than
  the velocity field fluctuations.}
\label{theta}
\end{figure}

Consider now the anisotropic evolution of ${\bf B}$ and ${\bf V}$
turbulent fluctuations in the Hall MHD regime of solar wind plasma,
i.e. the $kd_i > 1$ wavenumber band. Specifically we investigate the
degree of anisotropy associated with the nonlinear transfer of
turbulent energy in these fluctuations. The results of our 3D fully
compressible Hall MHD plasma simulations are plotted in \Fig{theta}.
In \Fig{theta}, we quantify the turbulent anisotropy associated with
small scale turbulent fluctuations that correspond to the rotational
${\bf B}$ and ${\bf V}$ fields to investigate which is more
anisotropic in the presence of the mean magnetic field. The evolution
of turbulent anisotropy, corresponding to small scale rotational ${\bf
  B}$ and ${\bf V}$ fluctuations, is determined from the following
relations,

\be 
\label{thetaB}
\theta_{\nabla \times {\bf B}}  = \tan^{-1} \left( \frac{\sum_{\bf k} |k_\perp|^2|i {\bf k} \times {\bf B}({\bf k},t)|^2}{\sum_k |k_\parallel|^2|i {\bf k} \times {\bf B}({\bf k},t)|^2 } \right)^{1/2}, \ee

\be 
\label{thetaV}
\theta_{\nabla \times {\bf V}}  = \tan^{-1} \left( \frac{\sum_{\bf k} |k_\perp|^2|i {\bf k} \times {\bf V}({\bf k},t)|^2}{\sum_k |k_\parallel|^2|i {\bf k} \times {\bf V}({\bf k},t)|^2 } \right)^{1/2}. \ee

For isotropy, $k_\perp \simeq k_\parallel$, hence the angle of
ansiotropy is close to $\theta \simeq 45^\circ$. Any deviation from
the isotropic spectrum can be seen if $\theta \neq 45^\circ$. It is
also clear from \eqs{thetaB}{thetaV} that $\theta > 45^\circ$
corresponds to the inertial range anisotropic turbulent spectrum that
is dominated by the cascades in the perpendicular direction such that
$k_\perp > k_\parallel$ whereas $\theta < 45^\circ$ describes $k_\perp
< k_\parallel$ cascades.  The evolution of the angular anisotropy
associated with small scale inertial range turbulent fluctuations
described by \eqs{thetaB}{thetaV} is depicted in \Fig{theta}. Several
important points emerge from \Fig{theta}.  Firstly, spectral
anisotropy associated with small scale inertial range turbulent
fluctuations in the velocity field is more pronounced than that in the
magnetic field fluctuations at large times. The angular evolution
associated with the anisotropic velocity field deviates markedly from
the isotropic angle $ \theta = 45^\circ$. On the other hand, magnetic
field fluctuations show a smaller degree of anisotropy compared to the
velocity field fluctuations. It then appears, in view of \Fig{aniso2},
that the velocity field spectrum is dominated by relatively large
scale turbulent fluctuations as compared to the magnetic field.

\begin{figure}
\vspace{200pt}
\begin{center}
\end{center}
\caption{\label{angle_mhd} Evolution of the degree of alignment in
  Hall MHD fluctuations. A progressive decrease in the angle of
  alignment from $90^\circ$ indicates the eventual weakening of ${\bf
    V} \times {\bf B}$ nonlinear interactions in the Alfv\'enic
  cascade regime of solar wind turbulence. }
\end{figure}

\section{Dynamical alignment  of turbulent fluctuations}
The nature and strength of nonlinear interactions resides critically
with their orientation with respect to each other in the presence of
an externally imposed or self-consistently generated large scale
magnetic field. For instance, the dominant nonlinear interactions in a
Hall MHD fluid are governed by the Lorentz force that contains a
nonlinear term corresponding to the ${\bf V} \times {\bf B}$
nonlinearity. Hence the orientation of the velocity field relative to
the magnetic field fluctuations plays a critical role in determining
the strength of nonlinear interactions.  This clearly means that
nonlinear interactions are dominated by those fluctuations that
possess the velocity field aligned perfectly orthogonal (at a
$90^\circ$ angle) to the magnetic field, i.e. ${\bf V} \perp {\bf
  B}$. The obliqueness, introduced primarily by deviation from the
orthogonality, tends to weaken the strength of nonlinear interactions
mediated by the ${\bf V} \times {\bf B}$ nonlinearity. The ${\bf V}
\times {\bf B}$ nonlinearity disappears (i.e. ${\bf V} \parallel {\bf
  B} = 0$) for fluctuations that carry parallel or anti-parallel
velocity and magnetic field fluctuations.

To understand the strength of the nonlinear interactions in Hall MHD
solar wind plasma, we determine the degree of alignment of the
velocity and magnetic field fluctuations by defining the following
alignment parameter (Podesta et al 2008) that spans the entire
$k$-spectrum in both the $kd_i>1$ (Hall MHD) and $kd_i<1$ (usual MHD)
regimes.
\[ \Theta(t) = \cos^{-1} \left(\frac{\sum_{\bf k} {\bf V}_{\bf k} (t) \cdot {\bf B}_{\bf k} (t)}{\sum_{\bf k}|{\bf V}_{\bf k}(t)| |{\bf B}_{\bf k}(t)|} \right). \]
The summation is determined from the modes by summing over the entire
spectrum. In this sense, the alignment parameter depicts an average
evolution of the alignment of velocity relative to the magnetic field
fluctuations. Note carefully that this alignment can vary locally from
smaller to larger scales, but the averaging (i.e. summing over the
entire spectrum) rules out any such possibility in our
simulations. Nonetheless, $\Theta$ defined as above enables us to
quantitatively measure the average alignment of the magnetic and
velocity field fluctuations while the nonlinear interactions evolve in
a turbulent solar wind plasma.

\begin{figure}
\vspace{200pt}
\begin{center}
\end{center}
\caption{\label{angle_hall} Evolution of the degree of alignment in
  the $kd_i>1$ regime of Hall MHD. Small scale fluctuations possess
  orthogonal velocity and magnetic fields and hence the angle of
  alignment stays close to $90^\circ$.}
\end{figure}

Unlike Podesta et al. (2008), where a probability distribution of
angular orientation corresponding to each mode is determined, we
follow the total (i.e. spectrally volume averaged) evolution of the
angle of alignment for the velocity and magnetic field fluctuations in
both the Alfv\'enic cascade (i.e. MHD) and kinetic Alfv\'en wave
(i.e. Hall MHD) regimes. Our results are plotted respectively in
\fig{angle_mhd}{angle_hall}. The Alfv\'enic cascade regime of MHD
turbulence $kd_i<1$ in the solar wind plasma possesses relatively
large scales ($kd_i<1$) in which the velocity and magnetic field
fluctuations are observed to be somewhat obliquely aligned. Hence our
simulations show that the angle of alignment evolves towards $\Theta <
90^\circ$, as depicted in \Fig{angle_mhd}.  Hence the strength of the
nonlinear interactions corresponding to the ${\bf V} \times {\bf B}$
nonlinearity is relatively weak. This result is to be contrasted with
characteristic turbulent length scales in the $kd_i>1$ regime. The
angle of alignment for the smaller scales corresponding to the
$kd_i>1$ regime is shown in \Fig{angle_hall}. Significant differences
are apparent in the angle of alignments associated with the large and
small scales \fig{angle_mhd}{angle_hall}.  It appears from our
simulations that the small scale fluctuations ($kd_i>1$) are nearly
orthogonal as seen in Fig (10). By contrast, the large scale
fluctuations ($kd_i<1$) in Fig (9) show a significant departure from
the orthogonality.

It is interesting to note that characteristic length scales that are
large compared to the ion inertial skin depth ($kd_i<1$) tend to
deviate from orthogonality (a tendency towards dynamic
alignment). Furthermore, these are the scales at which anisotropic
cascades are dominated by the higher rate of spectral transfer of
energy in the $k_\perp$ modes. This suggests the possibility that
anisotropic cascades are related to the orientation of the velocity
field relative to the magnetic field fluctuations. Our simulation
results described in Figs (\ref{aniso2}), (\ref{theta}),
(\ref{angle_mhd}) \& (\ref{angle_hall}) hint that the predominance of
spectral anisotropy at large scales and the obliqueness are related. A
heuristic and theoretical analysis of this relationship is beyond the
scope of this paper.

\section{Comparison between KAW and  whistler  spectra}
It is interesting to note the similarities and differences between the
characteristic Hall MHD and whistler wave turbulence spectra. Here we
discuss properties of whistler turbulence from the previous works by
\citep{dastgeer03,dastgeer05,dastgeer09,dastgeer10,biskamp} and compare
them with our present simulation results.  

While the two spectra correspond to essentially different frequencies
as depicted schematically in \Fig{fig1}, they are described by nearly
identical spectral power laws in the vicinity of high frequency KAW
modes (see the interface of boundary between III and IV in
\Fig{fig1}). Thus, the small scale KAW in the $kd_i>1$ regime and
relatively large scale whistler fluctuations exhibit a $k^{-7/3}$
spectrum. Although the slope of the inertial range spectrum for these
two modes is identical, the nonlinear physical processes yielding the
spectra are distinctively different.  Whistler modes are essentially
governed by the motion of electrons in a static neutralizing ion
background.  The characteristic frequencies therefore reside between
the ion and electron gyro frequencies ($\omega_{ci}\ll \omega \ll
\omega_{ce}$) and characteristic length scales lie between the ion
($d_i$) and electron ($d_e$) inertial lengths i.e., $d_i=c/\omega_{pi}
< \ell < d_e=c/\omega_{pe}$, where $\omega_{pi}$ and $\omega_{pe}$ are
respectively the ion and electron plasma frequency. By virtue of
$d_e$, there exist two inertial ranges that correspond to smaller
$kd_e>1$ and larger $kd_e<1$ lengthscales. Correspondingly, forward
cascade turbulent spectra in these regimes exhibit $k^{-5/3}$ and
$k^{-7/3}$ respectively as shown in schematic \Fig{fig1}. The whistler
spectrum $k^{-7/3}$ is produced essentially by fluctuations in
electron fluid while {\it ions are at rest}.  This is demonstrated in
2D \citep{biskamp,dastgeer05} as well as in 3D \cite{dastgeer10}
simulations.  As described in \cite{dastgeer10}, the whistler waves
cascade the inertial range turbulent energy typically through the
convective electron fluid velocity $v_e \sim \nabla \times{\bf
  B}$. Thus the typical velocity of the magnetic field eddy $B_{\ell}$
with a scale size $\ell$ can be represented by $v_e \simeq
B_{\ell}/\ell$. The eddy turn-over time is then given by \be \tau_{\rm
  whis} \sim \frac{\ell}{v_e} \sim \frac{\ell^2}{B_{\ell}}.
\label{whis-spectrum}
\ee 
It is this time scale that characterizes the nonlinear spectral
transfer of energy in fully developed whistler wave turbulence.
Kolmogorov-like dimensional arguments further yield a $k^{-7/3}$
spectrum in the $kd_e<1$ characteristic turbulent regime, as described
in \Fig{fig1}.  While the inertial range nonlinear cascades are
determined essentially by the eddy turn over or spectral transfer time
scale, the characteristic length scales longer and less than $d_e$ are
significantly influenced by the whistler interaction time scales in a
disparate manner \cite{dastgeer10}. By contrast, as described above,
the KAW spectrum mediated by Hall MHD processes in the $kd_i>1$ regime
is produced by the combined motion of electron and ion fluids amidst
density fluctuations that evolve on dispersive KAW time and length
scales.  The KAW fluctuations relax towards a $k^{-7/3}$ spectrum in
the $kd_i>1$ regime essentially by means of nonlinear mode coupling
interactions in which both the electron and ion fluid perturbations
participate on an equal footing. The most notable difference in the
energy cascade processes associated with the KAW and whistler modes
thus emerges from \Eqs{hmhd-spectrum}{whis-spectrum} where the energy
transfer rates are determined typically by different nonlinear fluid
velocities.

\section{Summary}
In summary, we have investigated the nonlinear and turbulent behavior
of a two fluid, compressible, three dimensional Hall MHD model.  Hall
MHD model is relevant to describe solar wind plasma and magnetospheric
and laboratory plasma environments. As exhibited by the solar wind
plasma, multiple scale characteristic fluctuations lead to a complex
power spectral density (or power spectrum) spanning very low
frequencies, corresponding to a flatter ($k^{-1}$) spectrum, followed
by the usual ``Alfv\'enic'' spectrum corresponding to a $k^{-5/3}$
spectrum. How the Alfv\'enic cascade is terminated in the solar wind
plasma, thus introducing a spectral discontinuity near the ion gyro
frequency, continues to challenge our understanding of nonlinear solar
wind turbulent processes near the $kd_i \sim 1$ band of
wavenumbers. The slope of inertial range beyond the $kd_i \sim 1$
wavenumbers varies between -2 and -5 (Smith et al 1990, Goldstein et
al 1994, Leamon et al 1999), depending upon what determines the
underlying nonlinear interactions.  Several mechanisms have been
proposed to describe the spectral discontinuity of turbulent energy
transfer in the solar wind plasma; these include kinetic Alfven waves
(KAWs) (Hasegawa 1976), electromagnetic ion-cyclotron-Alfv\'en (EMICA)
waves \cite{gary,yoon}, nonlinear Hall MHD interactions (Alexandrova
et al 2007, 2008; Shaikh \& Shukla 2008, 2008a), and suppression of
Alfv\'en fluctuations by proton cyclotron damping at intermediate
wavenumbers (Stawicki et al 2001), so exciting whistler waves which
are dispersive unlike Alfv\'en waves.  More complex interactions may
be expected due to the presence of highly obliquely propagating
un-damped kinetic Alfv\'en wave (with $\omega \ll \omega_{ci}$) as
proposed by \cite{howes08}.

In this paper, our simulation results make interesting non trivial
connections between regions that correspond to distinct wavenumbers in
the composite spectra sketched in \Fig{fig1}.  We find that the
spectra exhibit a break, consistent with the spacecraft observations
(Smith et al 1990, Goldstein et al 1994, Leamon et al 1999), in the
vicinity of the $kd_i \sim 1$ wavenumbers.  The characteristic
turbulent modes below these wavenumbers exhibit a $k^ {-5/3}$ like MHD
spectrum. By contrast, turbulent fluctuations scale as $k^ {-7/3}$ in
the inertial range characterized by $kd_i>1$.  The spectral index
$-7/3$ is mediated essentially by equipartition processes between the
turbulent velocity and magnetic field fluctuations. Note however that
turbulent equipartition, for instance $|B_k|^2 \simeq |V_k|^2$, is not
directly observed in the spacecraft data. Small scale inertial range
fluctuations in the $kd_i > 1$ regime tend to establish turbulent
equipartition amongst the characteristic modes in our simulations The
inertial range $k^ {-7/3}$ spectrum is a unique feature of nonlinearly
interacting multi-scale electromagnetic fluctuations that follows from
the turbulent equipartition of high frequency kinetic Alfv\'en modes.
We further find a considerable departure from turbulent equipartition
for the higher $k$ modes. The departure from equipartition in our
simulations can be ascribed essentially to compressibility, which we
find to be enhanced for the higher $k$ modes in the inertial range KAW
spectra.

  In the presence of a
large scale mean background magnetic field, small scale turbulent
fluctuations exhibit anisotropic cascades.  We find that the long
length scales in the $kd_i>1$ KAW regime are more anisotropic compared
to the shorter scales.  Dynamical alignment and angular distribution
of turbulence velocity and magnetic field fluctuations is found to
play a critical role in determining the degree of nonlinear
interactions.  We find that characteristic turbulent flucutations in
the $kd_i>1$ regime relax towards orthogonality, so that most of
turbulent scale fluctuations have velocity and magnetic fields that
are nearly orthogonal , i.e. ${\bf V} \perp {\bf B}$.  For large scale
fluctuations corresponding to the MHD regime, magnetic and velocity
fields are not perfectly orthogonal, being instead on average nearly
$70^\circ$ to each other. By contrast, small scale fluctuations in the
$kd_i>1$ KAW regime exhibit nearly perfect orthogonality in that the
average magnetic and velocity fields make an angle of nearly
$90^\circ$ with respect to each other. We finally noted the
similarities and differences between the KAW and whistler
spectra. Interestingly, the two modes exhibit a similar $k^{-7/3}$
spectrum in region III and IV of \Fig{fig1} that corresponds
respectively to the $kd_i>1$ and $kd_e<1$ regimes in wavenumber
space. While the KAW spectrum is dominated by the combined motion of
ions and electrons, the whistler wave spectrum is governed
predominantly by electron motion only in the presence of a static
neutralizing ion background.

The spectral properties of nonlinear Hall MHD are particularly
relevant for understanding the observed solar wind and heliospheric
turbulence.  Hall MHD may also be useful for understanding multi-scale
electromagnetic fluctuations and magnetic field reconnection in the
Earth's magnetosphere \cite{matt,phan} and in laboratory plasmas
\cite{porkolab,ono,yamada,carter}.

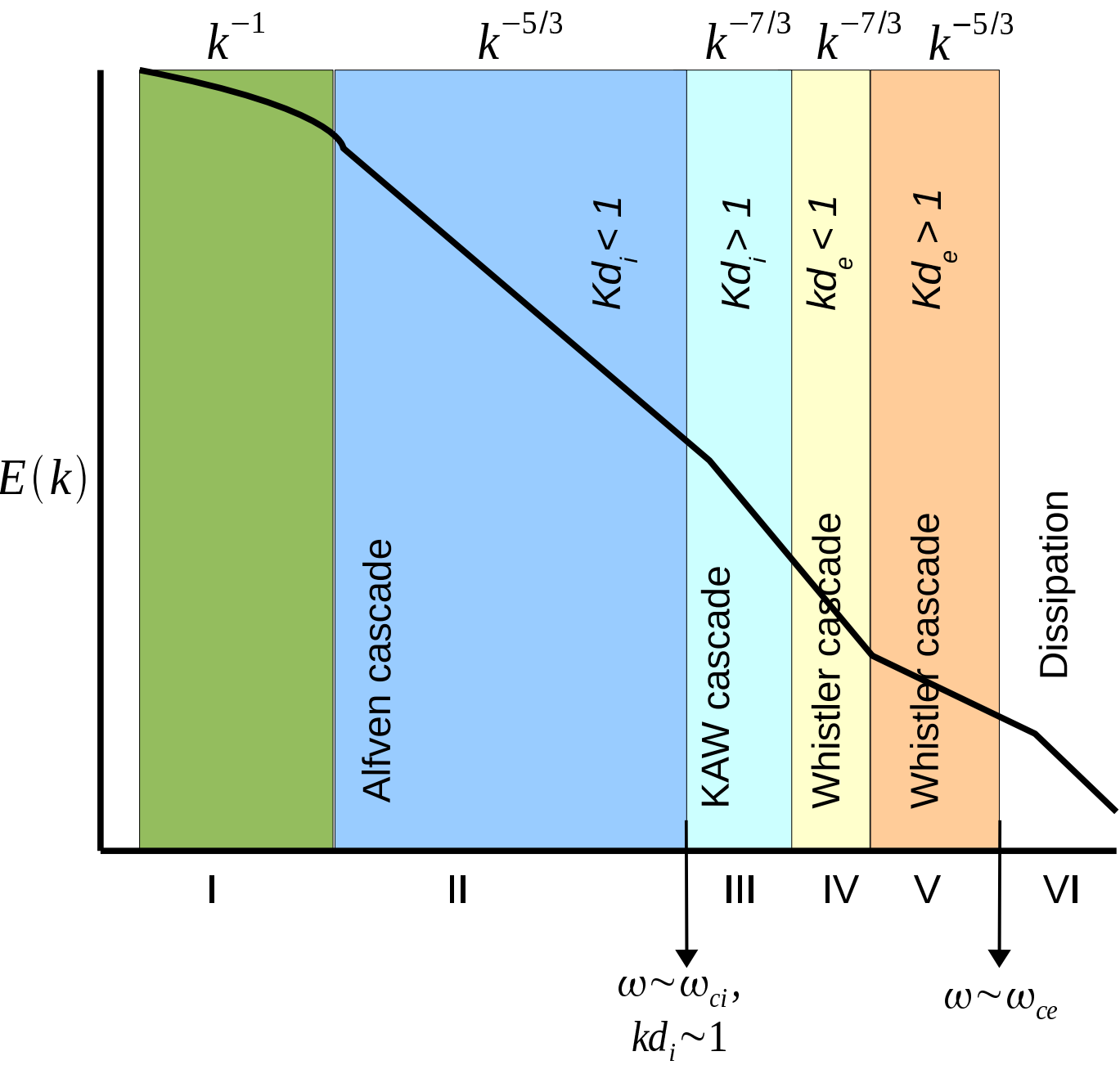
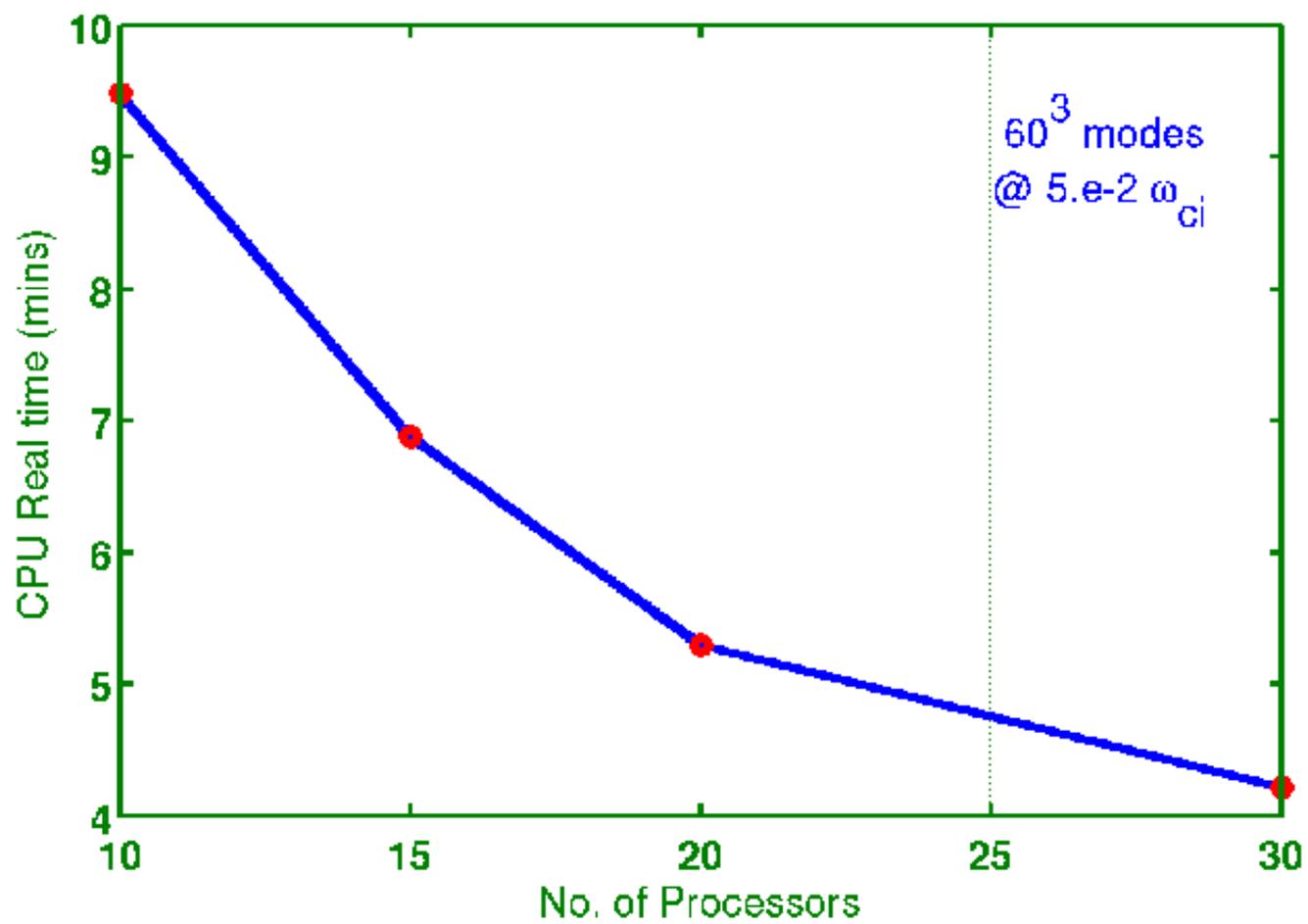
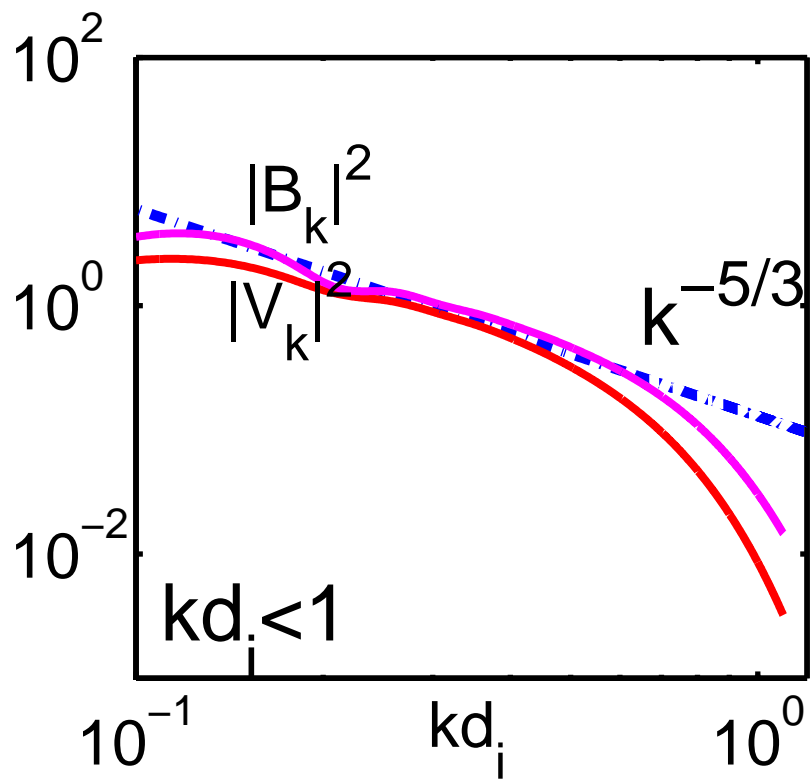
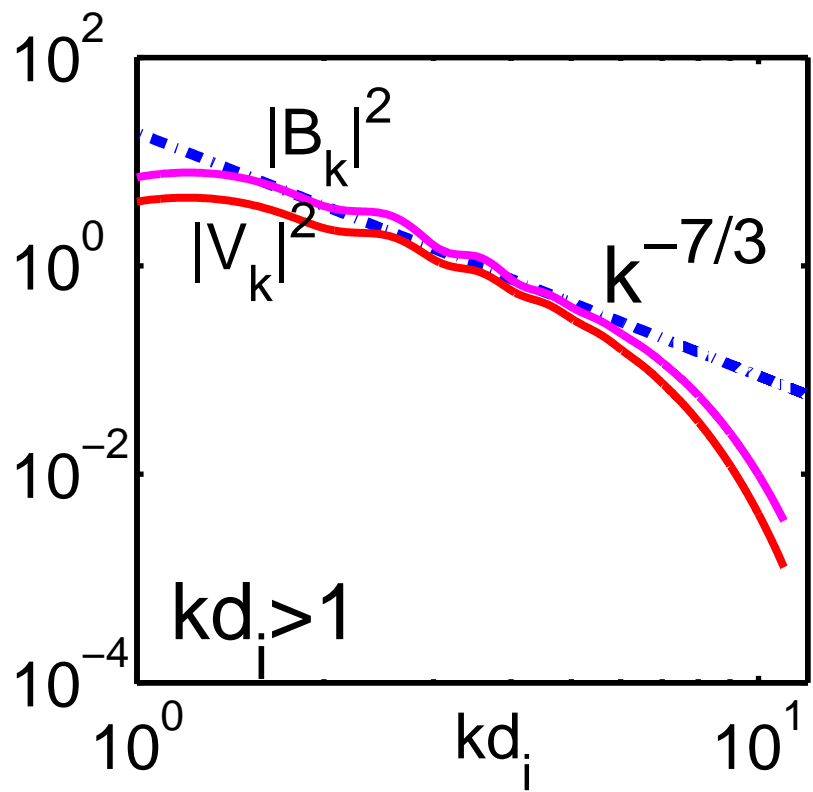
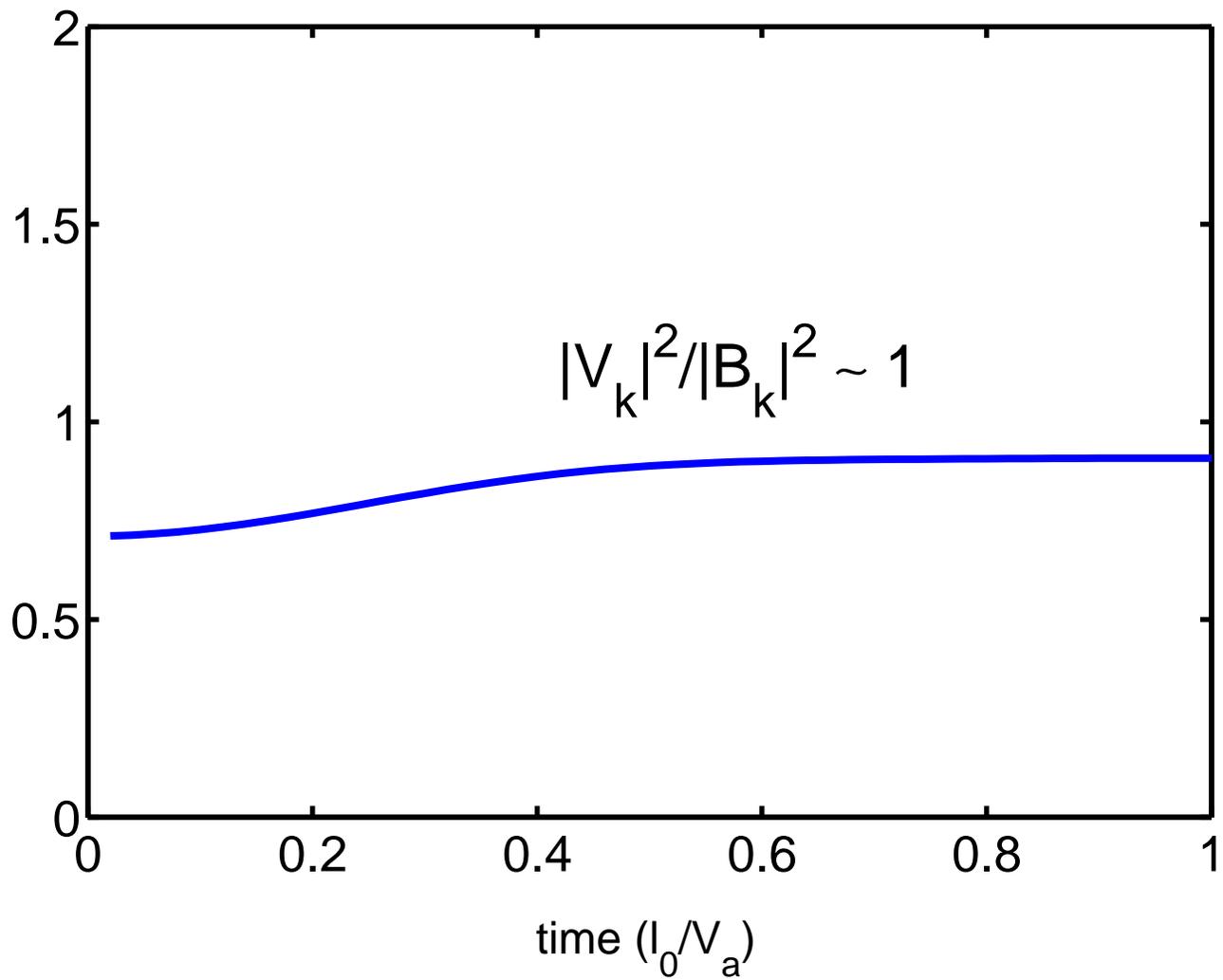
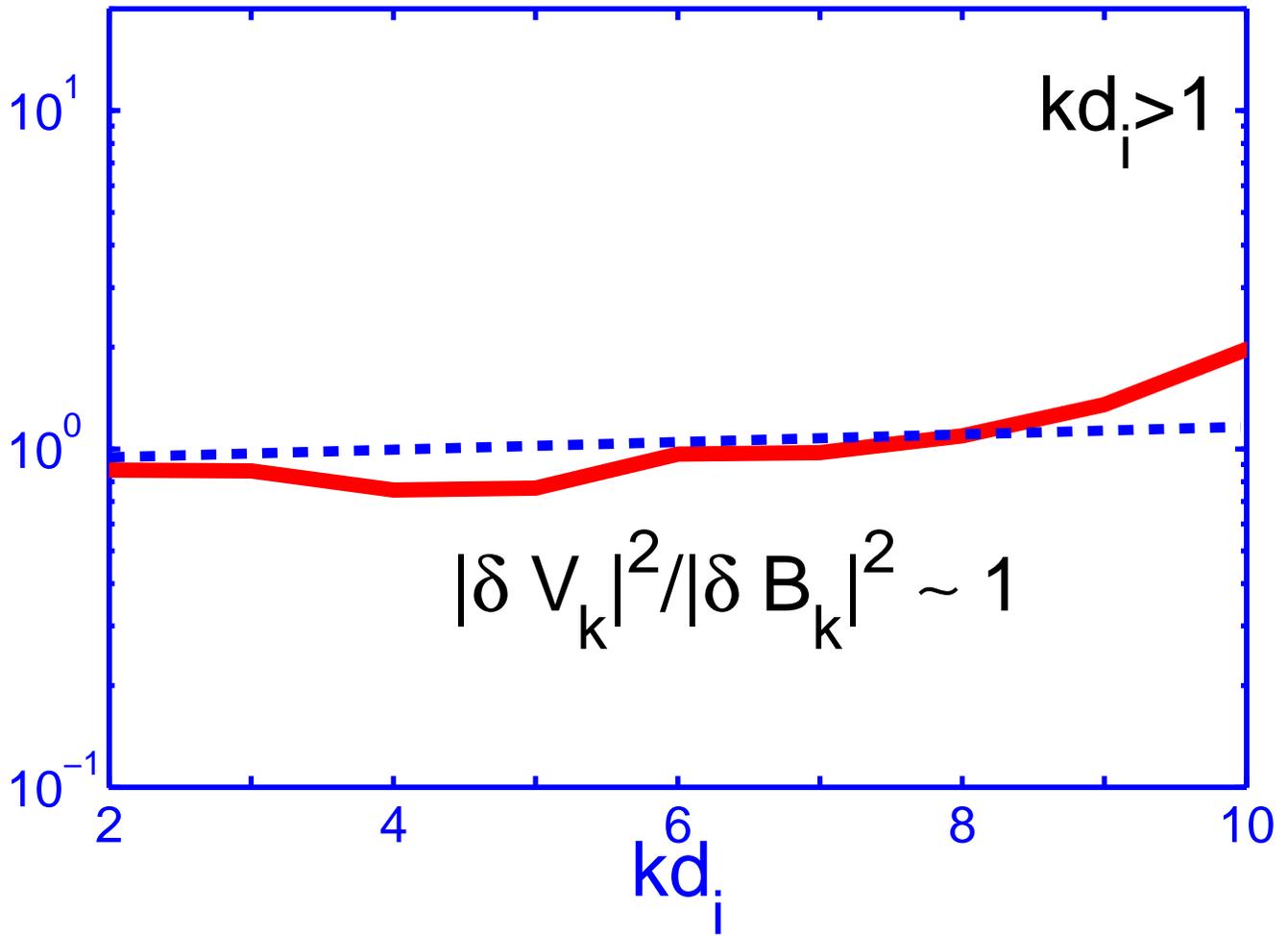
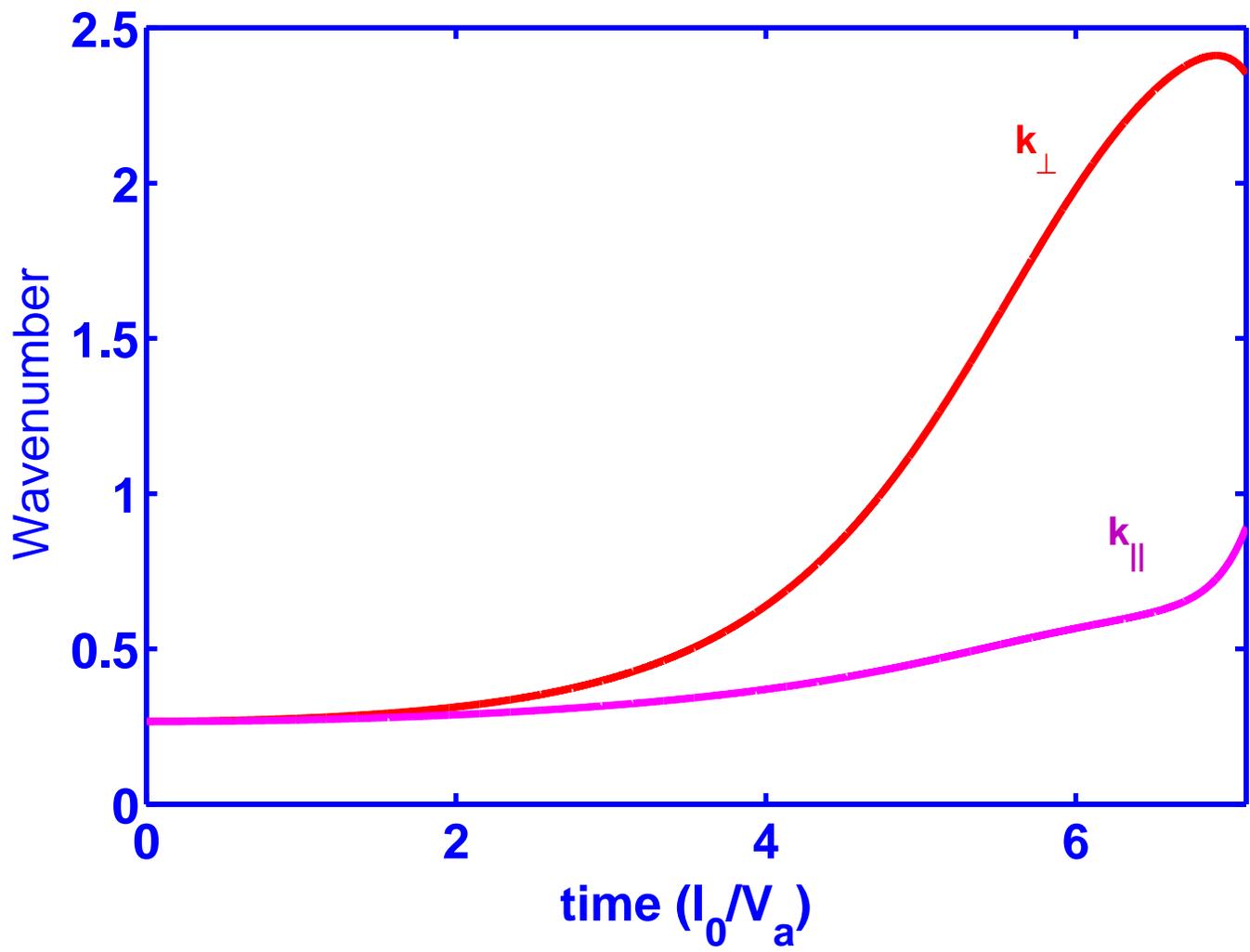
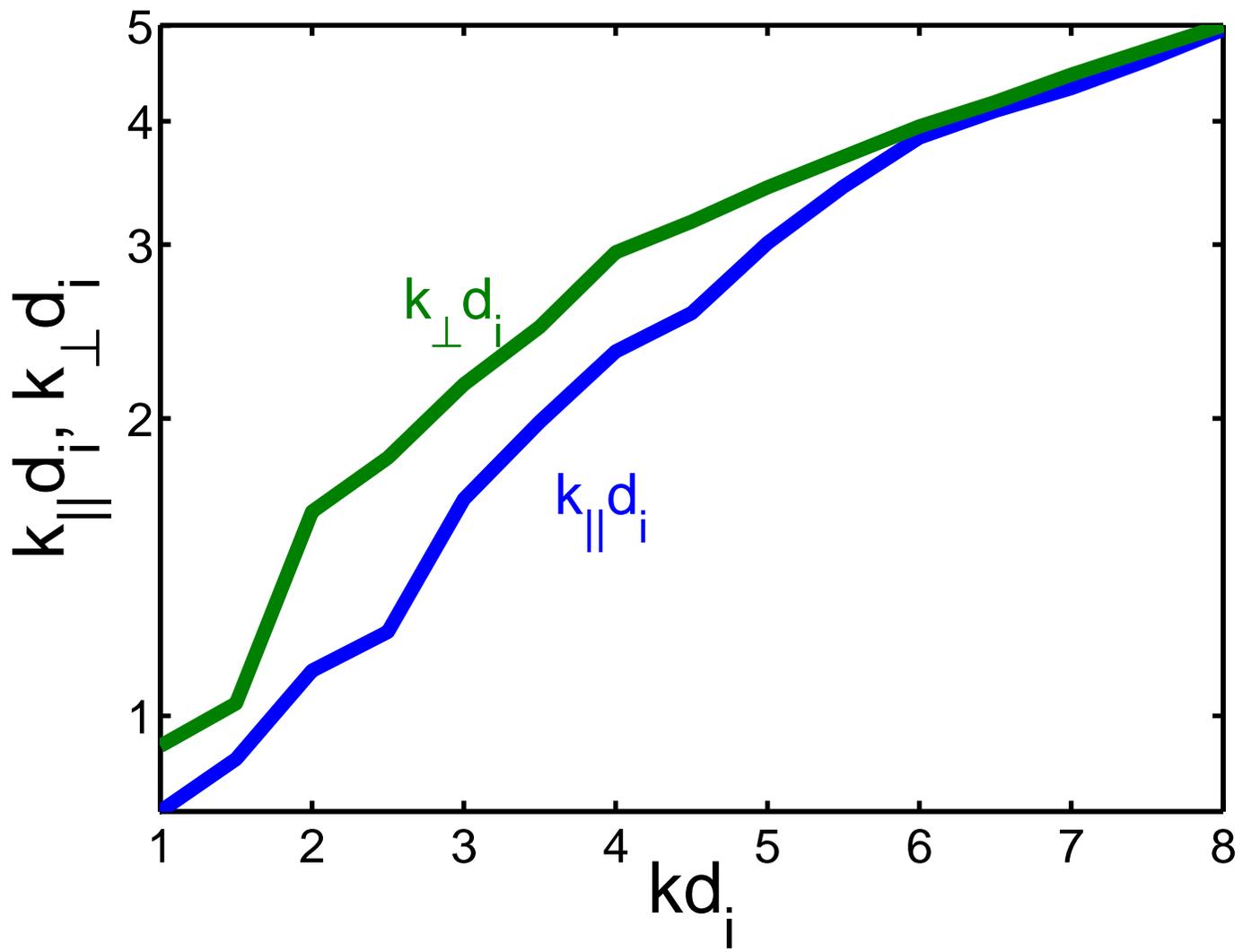
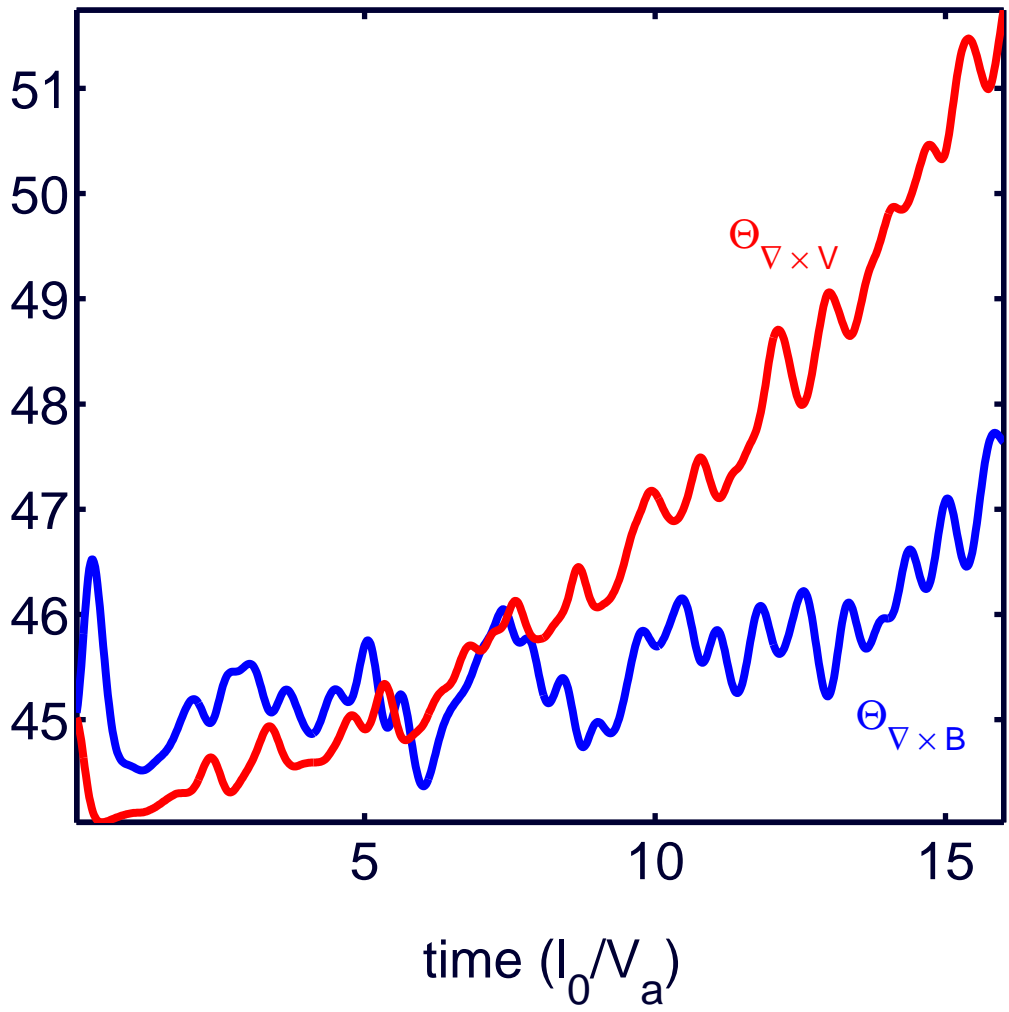
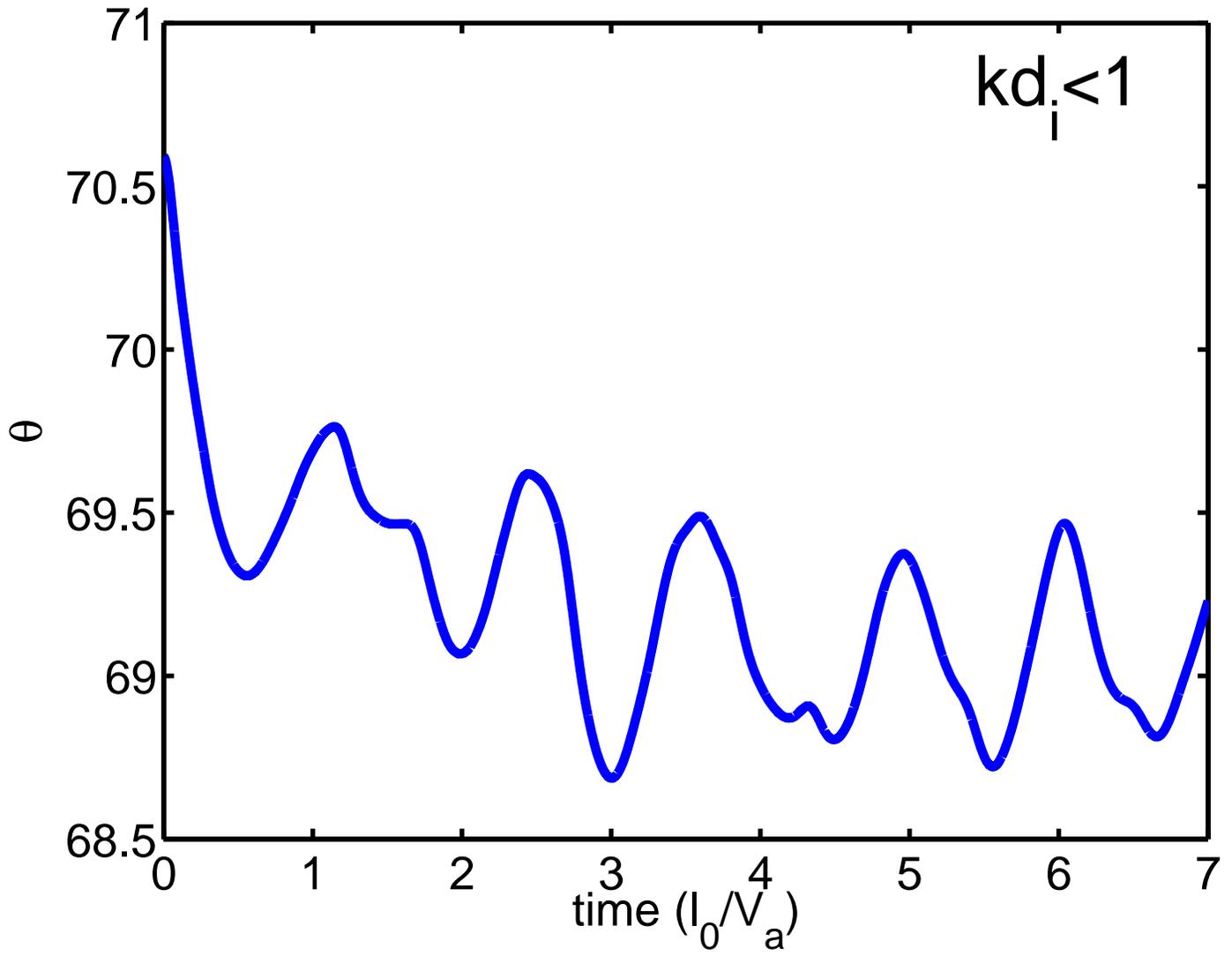
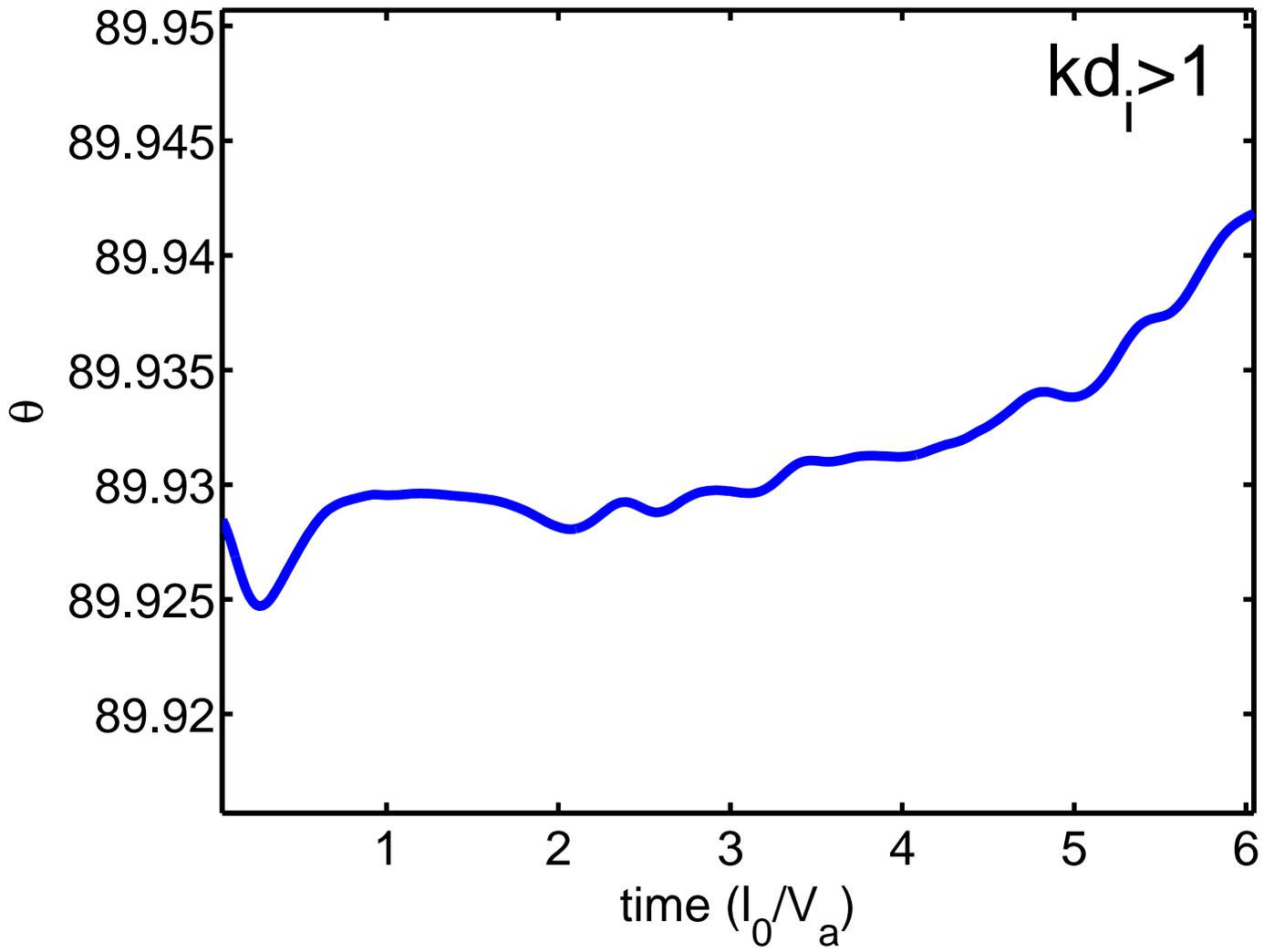
\end{document}